\documentclass[12pt]{iopart}
\usepackage{graphicx}
\usepackage{float}
\usepackage{adjustbox}
\expandafter\let\csname equation*\endcsname\relax
\expandafter\let\csname endequation*\endcsname\relax
\usepackage{amsmath}
\usepackage{multicol}
\usepackage{caption}

\DeclareRobustCommand{\rchi}{{\mathpalette\irchi\relax}}
\newcommand{\irchi}[2]{\raisebox{\depth}{$#1\chi$}}
\usepackage[bottom]{footmisc}

\begin{document}

\title[]{Bilayer h-BN Barriers for Tunneling Contacts in Fully-Encapsulated Monolayer $\mathbf{MoSe_2}$ Field-Effect Transistors}
\author{Talieh S. Ghiasi$^1$, Jorge Quereda$^1$, Bart J. van Wees$^1$}
\address{$^1$Zernike Institute for Advanced Materials, University of Groningen, Groningen, 9747 AG, The Netherlands}

\ead{t.s.ghiasi@rug.nl}

\begin{abstract}

The performance of electronic and spintronic devices based on two-dimensional semiconductors (2D SC) is largely dependent on the quality and resistance of the metal/SC electrical contacts, as well as preservation of the intrinsic properties of the SC channel. Direct Metal/SC interaction results in highly resistive contacts due to formation of large Schottky barriers and considerably affects the properties of the 2D SC. In this work, we address these two important issues in monolayer $\mathrm{MoSe_2}$ Field-Effect transistors (FETs). We encapsulate the $\mathrm{MoSe_2}$ channel with hexagonal Boron Nitride (h-BN), using bilayer h-BN at the metal/SC interface. The bilayer h-BN eliminates the metal/$\mathrm{MoSe_2}$ chemical interactions, preserves the electrical properties of $\mathrm{MoSe_2}$ and reduces the contact resistances by prevention of Fermi-level pinning. We investigate electrical transport in the monolayer $\mathrm{MoSe_2}$ FETs that yields close to intrinsic electron mobilities ($\approx 26\ \mathrm{cm^2 V^{-1} s^{-1}}$) even at room temperature. Moreover,
we experimentally study the charge transport through Metal/h-BN/$\mathrm{MoSe_2}$ tunnel contacts and we explicitly show that the dielectric bilayer of h-BN provides highly efficient gating (tuning the Fermi energy) of the $\mathrm{MoSe_2}$ channel at the contact regions even with small biases. Also we provide a theoretical model that allows to understand and reproduce the experimental $I-V$ characteristics of the contacts. These observations give an insight into the electrical behavior of the metal/h-BN/2D SC heterostructure and introduce bilayer h-BN as a suitable choice for high quality  tunneling contacts that allows for low energy charge and spin transport.    

\end{abstract}
\\

\maketitle
\ioptwocol

\section{Introduction}
\begin{figure*}[ht]
\centering\includegraphics[width=\textwidth]{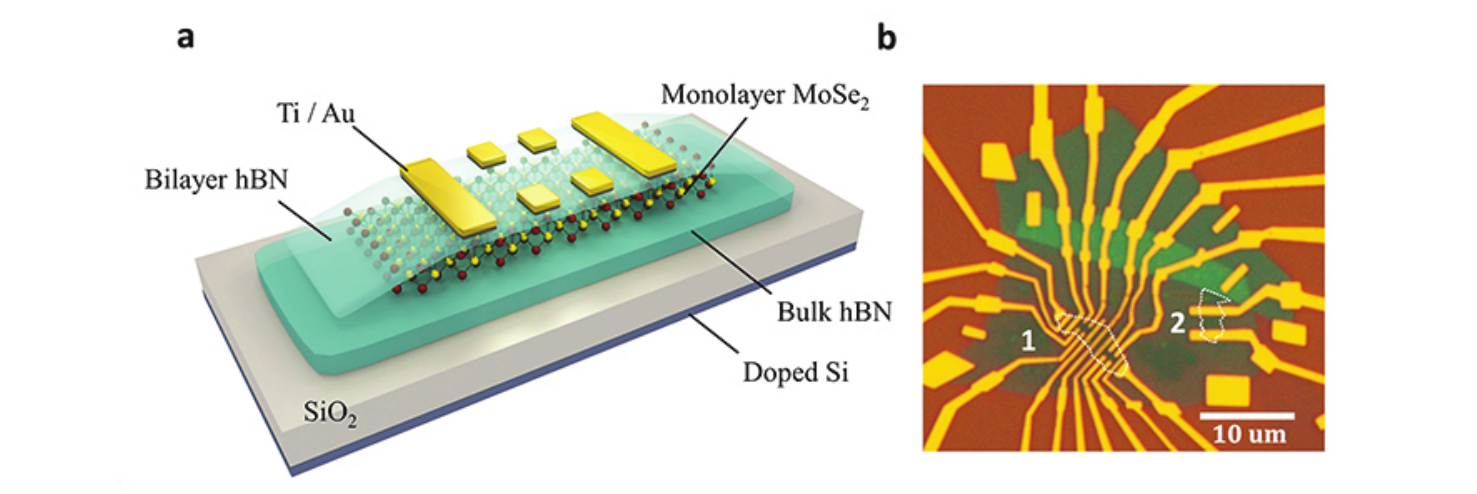}
\caption{\textbf{Device geometry.} (a) Sketch of the BN-encapsulated 1L $\mathrm{MoSe_2}$ FET with Ti/Au contacts. (b) Optical microscope image of the fabricated device. The edges of the separate $\mathrm{MoSe_2}$ flakes are shown by the white dashed lines. In device 1 and 2, the $\mathrm{MoSe_2}$ flakes are covered with 2-layer and 5-layer h-BN, respectively.}
    \label{fig:one}
\end{figure*}

Atomically thin Transition Metal Dichalcogenides (TMDs) are among the most promising materials for nano-electronics. In particular, single-layer TMDs can be used as a high-mobility semiconductor channel in field effect transistors (FETs), yielding significant on/off current ratios ($I_\mathrm{on}/I_\mathrm{off}>10^8$) \cite{radisavljevic2011single} and reduced power dissipation \cite{schwierz2010graphene}. In recent years, the coupled spin-valley physics in TMDs \cite{xiao2012coupled}  has also attracted broad attention, since it provides a new opportunity for (opto-) spin-valleytronic applications \cite{mak2014valley,sanchez2016valley}.

A major challenge for efficient charge and spin transport in TMD-based FETs is to achieve high quality, low resistive electrical contacts at the source and drain electrodes \cite{das2012high, chen2013control, guo2014study, cui2017low, avsar2017van, doi:10.1021/nl4010157, doi:10.1021/acsami.5b08559,dankert2014high, lee2015statistical, wang2016high}. Direct Metal/TMD interaction causes Fermi-level pinning and formation of large Schottky barriers and affects the electrical properties of the 2D TMD layer \cite{allain2015electrical, li2014thickness}. Also, the atomic thickness of the 2D channel falls below the width of the charge depletion region at the Metal/TMD contact. These facts lead to highly resistive electrical contacts that significantly limits the charge injection/detection efficiency in these FETs.

An effective strategy to address these issues is to use an insertion layer in between the TMD and metal, such as MgO, TiO$_2$, $\mathrm{Ta_2O_5}$ and h-BN \cite{doi:10.1021/nl4010157, doi:10.1021/acsami.5b08559,dankert2014high, lee2015statistical, wang2016high, cui2017low, avsar2017van}. Among them, h-BN is more promising for 2D contacts with TMDs since deposition of other mentioned materials on the dangling bond-free surface of TMD can lead to formation of isolated atomic islands and therefore poor quality of the contacts. Further, it was recently shown that bilayer h-BN tunnel barriers are highly efficient for spin injection \cite{gurram2017bias}. However, to our knowledge, a detailed description of the electrical behavior of these contacts for monolayer $\mathrm{MoSe_2}$ FETs is yet to be provided. In this work, we use an exfoliated bilayer h-BN as a tunnel barrier between the $\mathrm{MoSe_2}$ channel and the Ti electrodes that disrupts the Metal/TMD chemical interactions \cite{farmanbar2015controlling, farmanbar2016ohmic}. Also, as recently reported \cite{bokdam2014schottky}, the chemisorption of h-BN to Ti, lowers the Ti workfunction by 0.78 eV, improving its band alignment with the $\mathrm{MoSe_2}$ conduction band. 

Using a theoretical model for the electrical behavior of the metal/h-BN/TMD contacts, we are able to describe and reproduce the experimental observations in the three-terminal $I-V$ characteristics. In particular, we show that the strong out-of-plane electric fields formed at the contact interface can considerably modulate the doping of the underlying channel. 

We also evaluate the electrical performance of the monolayer $\mathrm{MoSe_2}$. The charge transport studies to date on monolayer $\mathrm{MoSe_2}$ FETs are limited to two-terminal measurements on non-encapsulated channels \cite{wang2014chemical,chang2014monolayer}. However, in this work, the use of metal/h-BN electrodes, instead of direct metal/TMD contacts, allows for full h-BN encapsulation of the $\mathrm{MoSe_2}$ channel that reduces the effect of (impurity dependent) Coulomb and roughness scatterings \cite{dean2010boron, doi:10.1021/acs.nanolett.6b02788} and prevents degradation of the $\mathrm{MoSe_2}$ crystal \cite{ahn2016prevention}. As further discussed below, the FETs are fabricated using a contact geometry that reduces the interaction between the electrodes and the channel. Thus, we can expect our four-terminal measurements to reflect the intrinsic behavior of the monolayer $\mathrm{MoSe_2}$ at room temperature.

Figure \ref{fig:one} shows a sketch and an optical microscope image of the studied FETs. In these devices the monolayer $\mathrm{MoSe_2}$ channel is encapsulated between bilayer and bulk h-BN as the top and bottom flakes, respectively. The h-BN/$\mathrm{MoSe_2}$/h-BN heterostructure is stacked on a $\mathrm{SiO_2}$(300 nm)/doped Si substrate, using a dry pick up technique \cite{Zomer} which provides polymer-free interfaces. The Ti (5 nm)/Au (75 nm) electrodes are fabricated by e-beam lithography (EBL), followed by e-beam evaporation of the metals at UHV (see Methods for details). The choice of Ti for the electrodes in these devices is due to its low work-function (4.33 eV) which closely matches the electron affinity of a monolayer $\mathrm{MoSe_2}$ ($E_\mathrm{{C}}$ = 3.99 eV) \cite{gong2013band}. 

Our sample contains two devices that are numbered in the optical image of Figure \ref{fig:one}b. Device 1 and 2 address the separate 1L-$\mathrm{MoSe_2}$ flakes which are covered with 2- and 5-layer h-BN, respectively. A homogeneous bulk h-BN flake is used as bottom layer in both devices and the doped Si layer is used as a back-gate electrode.

\section{Two-terminal measurements}

\begin{figure*}[ht]
\centering\includegraphics[width=\textwidth]{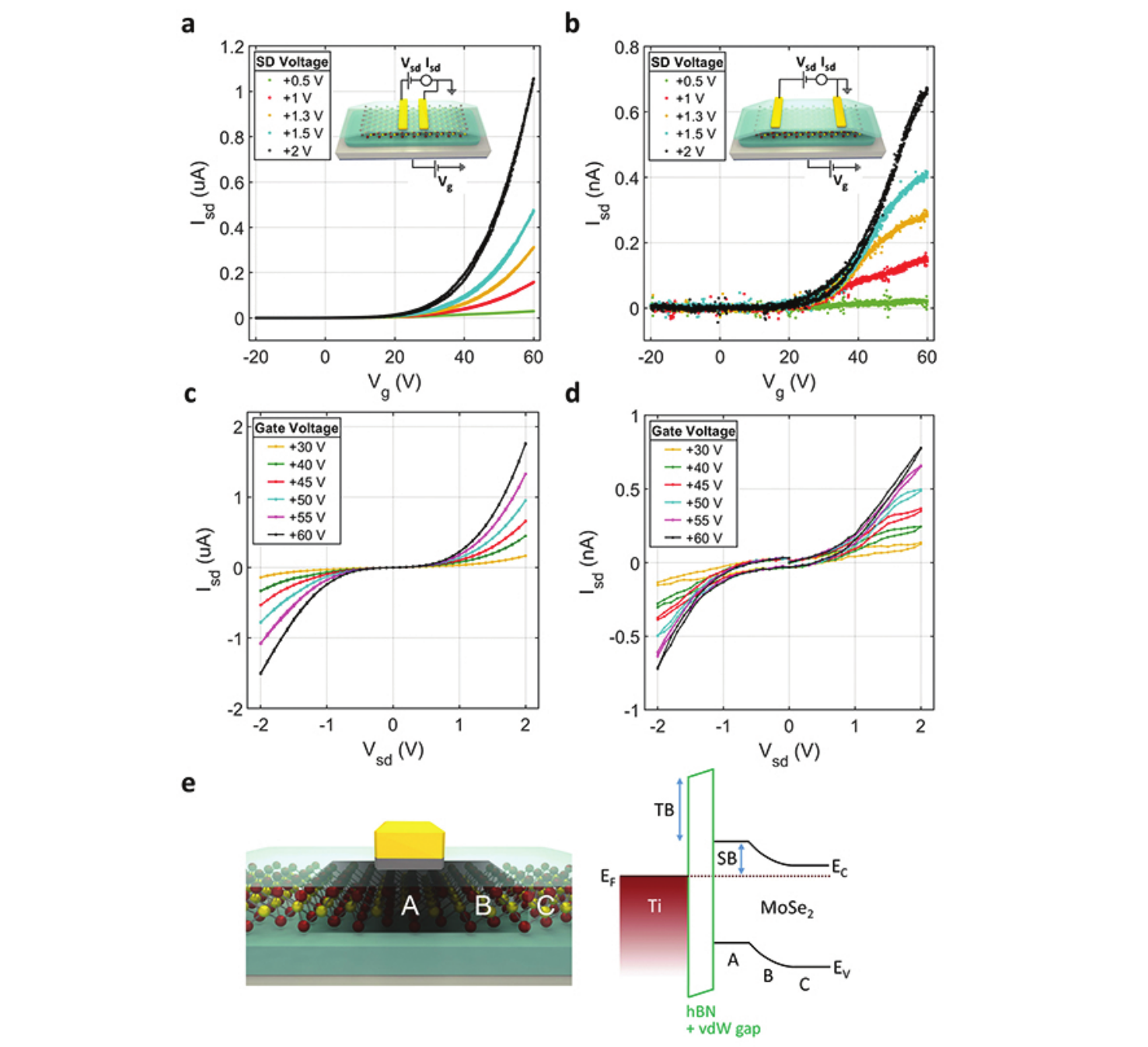}
\caption{\textbf{Two-terminal electrical measurements.} (a,b) Source-drain (SD) current ($I_\mathrm{{sd}}$) as a function of gate voltage ($V_\mathrm{{g}}$) measured at different SD biases ($V_\mathrm{{sd}}$) on (a) 1L-$\mathrm{MoSe_2}$ (channel length of $L_\mathrm{{ch}}$= 0.93 um and channel width of $W_\mathrm{{ch}}$= 2.8 um), covered with 2 layers of h-BN and (b) 1L-$\mathrm{MoSe_2}$ (with $L_\mathrm{{ch}}$= 2.54 um and $W_\mathrm{{ch}}$= 1.6 um), covered with 5 layers of h-BN. Sketches of the device and measurement geometry are shown in the insets. (c,d) SD $I-V$ characteristics at different $V_\mathrm{{g}}$ for the FETs with 2L (c) and 5L (d) h-BN. (e) Schematic drawing and the corresponding band diagram of a Ti/h-BN/$\mathrm{MoSe_2}$ heterostructure. The black gradient below the contact illustrates the highly depleted region (A) underneath the contact that extends laterally in the 2D SC through region (B) until it reaches the non-depleted region (C).}
\label{fig:two}
\end{figure*}

We start with evaluation of the FET performance by two-terminal (2T) measurements. All the measurements are done at room temperature and in vacuum ($<10^{-4}$ mbar). As shown in the device sketches of Figure \ref{fig:two}, the SC channel is connected to the source and drain electrodes through the bilayer h-BN and is separated from the gate electrode by the bulk h-BN and $\mathrm{SiO_2}$. The source electrode is grounded and a voltage ($V_\mathrm{{sd}}$) is applied on the drain electrode, while measuring the source-drain current ($I_\mathrm{{sd}}$).  We measure $I_\mathrm{{sd}}$ as a function of gate voltage ($V_\mathrm{{g}}$) (transfer curve) at different $V_\mathrm{{sd}}$ for the both devices of 1 and 2 (Figure \ref{fig:two}a and b) that shows gate-modulation of channel conductivity by tuning the density of charge carriers in the $\mathrm{MoSe_2}$ channel. 

The transfer curves show an almost hysteresis-free n-type behavior, indicating that the Fermi level of the SC is closer to the conduction band. Therefore $I_\mathrm{{sd}}$ increases for $V_\mathrm{g}$ larger than the threshold voltage ($V_\mathrm{{th}}$), as the Fermi energy approaches the conduction band minimum ($E_\mathrm{{C}}$). Comparison of the 2T measurements performed in device 1 and 2 shows that the 2T current injected through 2L h-BN is much larger than the current injected through 5L h-BN. From this 2T current ratio and square resistance of the channel (measured in four-terminal geometry, see next section), we estimate that the resistance of the contacts with 5L h-BN is 4 orders of magnitude larger than the ones with 2L h-BN. This observation is consistent with the fact that the increase in the number of h-BN layers reduces the charge tunneling efficiency into the channel \cite{britnell} but does not show considerable improvement in lowering the Schottky barrier height \cite{wang2016high}.  

The 2T source-drain $I-V$ measurements (Figure \ref{fig:two}c and d), show the non-linear, almost symmetric behavior, indicating the dominance of the contacts in the 2T charge transport (otherwise we would expect linear $I-V$ characteristics). In the SD $I-V$ curves of device 2 (Figure \ref{fig:two}d), we observe that $I_\mathrm{{sd}}$ starts to saturate at large $V_\mathrm{{sd}}$. This saturation can be attributed to the ``pinch-off" region in the SC channel, because the applied $V_\mathrm{{sd}}$ acts as a gate and can create a depletion region in the channel close to the drain contact \cite{ng2002complete}. The saturation of the current at such low SD biases shows that these contacts can considerably gate the channel. The $V_\mathrm{{sd}}$ at which the saturation of current happens, depends on the density of charge carriers in the channel. The higher $V_\mathrm{{g}}$, induces a larger density of charge carriers and therefore charge depletion close to the contacts happens at larger $V_\mathrm{{sd}}$. 

Therefore, the electrical performance of a two-terminal TMD-based FET is highly dependent on the electrical response of the contacts. When the metal comes in vicinity of the n-type 2D SC, the SC gets charge-depleted \cite{allain2015electrical}. In Figure \ref{fig:two}e, we show the schematic drawings of the metal/h-BN/SC interface, where the black gradient illustrates the depleted region that is gradually disappearing in the lateral direction of the SC along the channel. The corresponding band diagram shows the only energy level available for the states in the 2D SC channel beneath the contact (region A) which reaches the less depleted region close to the edge of the contact (region B) and laterally expands into the SC where there will be no charge depletion and the SC bands preserve their intrinsic energy levels (region C) \cite{allain2015electrical}. In this situation the charge carriers that are injected far from the edge of the contact, firstly encounter region A before reaching region B. The low density of charge carriers at these depleted regions makes the contact areas highly resistive in comparison with the channel. Therefore, in the 2T measurements, modulation of the $I_\mathrm{{sd}}$ as a function of $V_\mathrm{{sd}}$ and $V_\mathrm{g}$ is considerably affected by the contact regions that have the main contribution in the voltage drop in the 2T circuit. In the band diagram of Figure \ref{fig:two}e, we also show the tunnel barrier formed by the van der Waals (vdW) gap in addition to the h-BN layer. This gap might not be present in the direct metal/TMD interface, since the interaction can be accompanied with covalent bonding (strong hybridization)\cite{kang2014computational}. 

We also compare the performance of the 2T FETs, fabricated with encapsulated and non-encapsulated 1L-$\mathrm{MoSe_2}$ channels (see Supporting Information (SI), section 3). We observe that the BN-encapsulation of the channel considerably improves the 2T electrical transport in the 1L-$\mathrm{MoSe_2}$. This is because TMDs are highly sensitive to environmental adsorbates or scattering centers \cite{late2012hysteresis}. This fact, in addition to the formation of large SB at the direct Metal/TMD interface leads to one order of magnitude lower 2T conductivity and mobility in the non-encapsulated $\mathrm{MoSe_2}$ channel. Also there is a large hysteresis in the electrical measurements on non-encapsulated $\mathrm{MoSe_2}$ FET (about 29 V), while the BN-encapsulated sample shows almost no-hysteresis for the same gate bias sweeping rate.

\section{Four-terminal measurements}

\begin{figure*}[ht]
\centering\includegraphics[width=\textwidth]{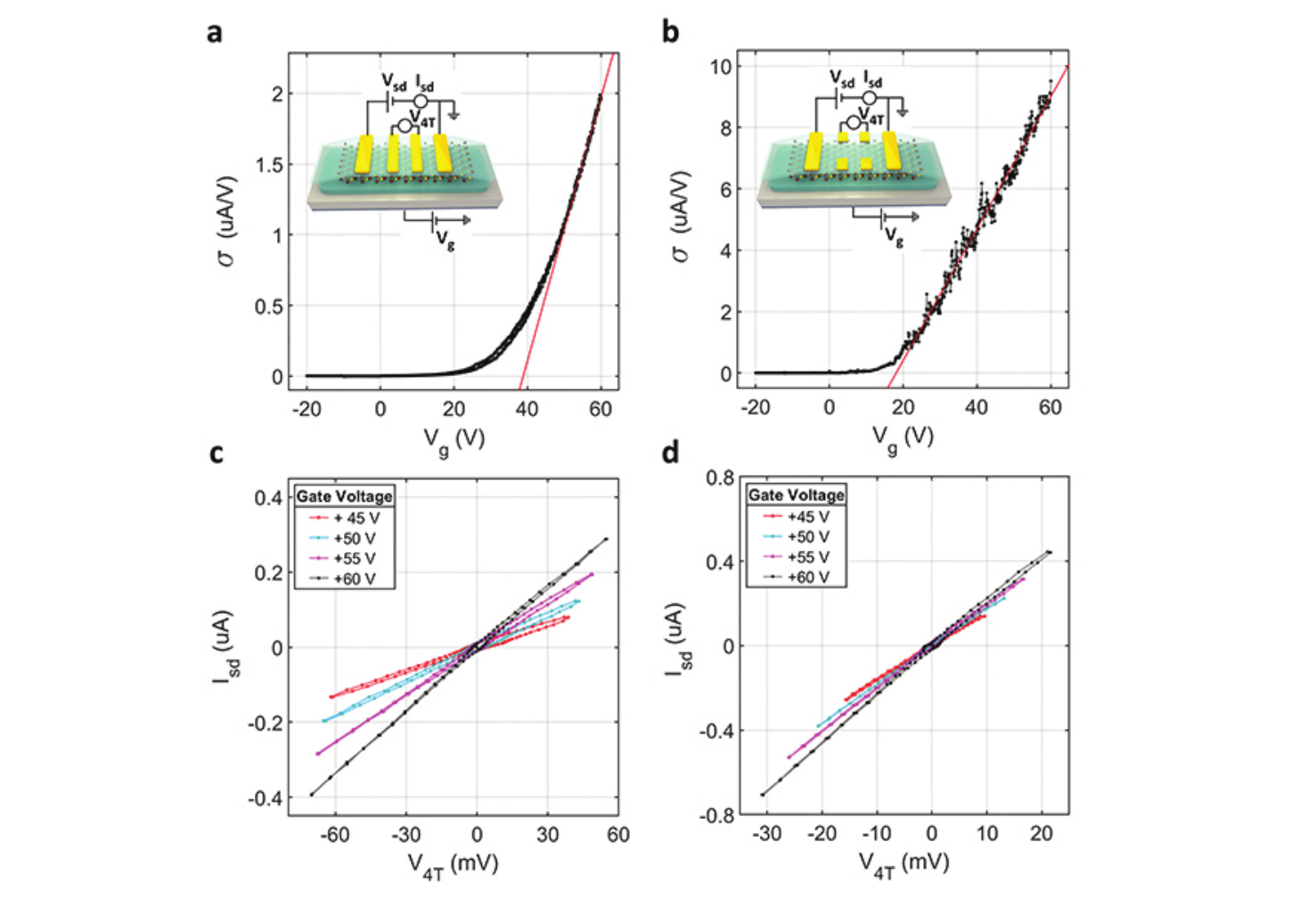}
\caption{\textbf{Four-terminal electrical characterization of the BN-encapsulated $\mathbf{MoSe_2}$ channel.} (a,b) Channel conductivity ($\sigma$) for the device with bilayer h-BN tunnel barrier as a function of $V_\mathrm{{g}}$, where the four-terminal voltage ($V_\mathrm{{4T}}$) was measured using (a) electrodes that fully cover the channel (crossing electrodes) and (b) using side electrodes. The insets are sketches of the device and measurement geometries. (c,d) four-terminal $I-V$ characteristics at different gate voltages ($V_\mathrm{{g}}$) as measured with the crossing (c) and the side (d) electrodes.}
\label{fig:three}
\end{figure*}

In order to investigate the intrinsic electrical properties of the monolayer $\mathrm{MoSe_2}$ channel, we perform four-terminal (4T) measurements. As shown in the device sketches of Figure \ref{fig:three}, we apply the $V_\mathrm{{sd}}$ to the outer electrodes and measure $I_\mathrm{{sd}}$, while probing the voltage drop across the inner electrodes ($V_\mathrm{{4T}}$). First, we perform these measurements with the electrodes that are crossing the full width of the channel (crossing electrodes). Figure \ref{fig:three}a shows the dependence of channel conductivity ($\sigma = I_\mathrm{{sd}}L_\mathrm{{ch}} V_\mathrm{{4T}}^{-1}W_\mathrm{{ch}}^{-1}$) on $\mathrm{V_{g}}$ and Figure \ref{fig:three}c shows the linear dependence of $V_\mathrm{{4T}}$ on $I_\mathrm{{sd}}$. From the slope of the SD $I-V$ curves, we extract the square resistance ($\mathrm{R_{sq}}$) of the channel at different $\mathrm{V_{g}}$ ($\mathrm{R_{sq}}$ = 0.5 M$\mathrm{\Omega}$ to 1.5 M$\mathrm{\Omega}$ for the range of $\mathrm{V_{g}}$ = 60 V to 45 V). 

However, the depleted regions of the contact areas (shown in Figure \ref{fig:two}c) can contribute to the 4T measurements, performed by using the crossing contacts. In order to avoid the effect of the depleted regions and address the intrinsic electrical behavior of the channel material, we perform the 4T measurements using side contacts (the electrodes that are partially covering the channel). The minimized overlap of the side contacts with the $\mathrm{MoSe_2}$ flake diminishes screening of the gate-induced charges in the channel. Figure \ref{fig:three}b shows the measurement geometry and gate dependency of the channel conductivity. The dependence of $\mathrm{V_{4T}}$ as a function of $\mathrm{I_{sd}}$ for different $\mathrm{V_{g}}$ is shown in Figure \ref{fig:three}d. The considerable difference between the results of these two 4T measurement geometries (Figure \ref{fig:three}a and b) makes it clear that the role of the mentioned depleted regions underneath the contacts should not be overlooked. The conductivity of the channel, derived from the 4T measurements with the side contacts, clearly shows linear behavior for $\mathrm{V_{g}}>\mathrm{V_{th}}= 20\ \text{V}$, while the conductivity measurements with the crossing contacts are behaving more similar to the 2T measurements (Figure \ref{fig:two}a) that show gradual increase versus $\mathrm{V_{g}}$ and hardly reaches the linear regime. By the linear fit to the conductivity curve for $\mathrm{V_{g}}>\mathrm{V_{th}}$ (in \ref{fig:three}b), we extract the electron field-effect mobility of $\mathrm{\mu_{FE}}\approx 26\ \mathrm{cm^2 V^{-1} s^{-1}}$, considering the effective width of the channel since it is partially covered by the side contacts. The extracted $\mathrm{\mu_{FE}}$ in the 1L-$\mathrm{MoSe_2}$ channel is close to the intrinsic value predicted by theoretical studies \cite{jin2014intrinsic}. This indicates that the electrical properties of the channel are well preserved in our sample. The BN-encapsulation of the channel reduces the Coulomb scatterings caused by charged impurities at the interfaces or on the channel surface (e.g. fabrication residues) and also eliminates roughness scatterings originating from the Si substrate \cite{quereda2014single,wang2015electronic} that could highly affect the charge carrier mobility \cite{ando1982electronic, chen2008intrinsic}. From the linear $I-V$ curves measured with the side contacts, we extract the values of $\mathrm{R_{sq}}$= 130 k$\mathrm{\Omega}$ to 190 k$\mathrm{\Omega}$ for the range of $\mathrm{V_{g}}$ = 60 V to 45 V. The channel resistances is about 1 order of magnitude smaller than the ones measured with the crossing contacts. The derived $\mathrm{R_{sq}}$ and $\mathrm{\mu_{FE}}$ are consistent with the results of the same measurements done with the other pairs of side contacts (see Figure \ref{fig:one}b). In contrast, the 4T measurements performed with the crossing electrodes vary for different contacts along the channel because of the inhomogeneity of the contact regions (can be due to the presence of bubbles at the 2D interfaces or variation of the contact area). This observation is another evidence for contribution of the depletion regions from the crossing contacts in the latter case.\

\section{Modeling of metal/h-BN/TMD tunnel contact and three-terminal measurements}

\begin{figure*} [ht]
    \centering
    \includegraphics[width=\textwidth]{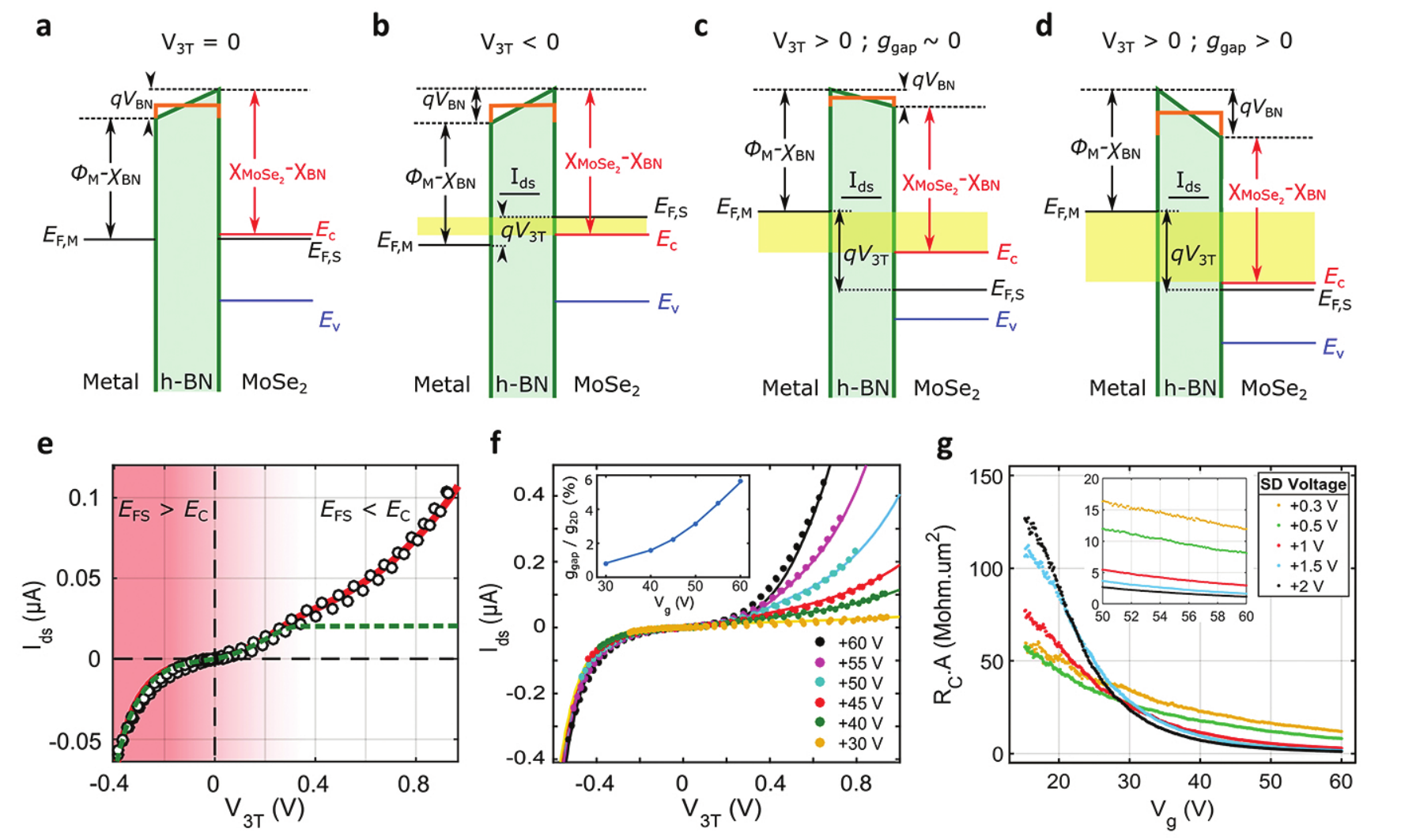}
   \caption{\textbf{Model for the Metal/h-BN/TMD tunnel contact.} (a-d) Band diagrams of the tunnel contact for (a) $V_\mathrm{3T} = 0$, (b) $V_\mathrm{3T} < 0$, (c) $V_\mathrm{3T} > 0$ and $g_\text{gap} = 0$ and (d) $V_\mathrm{3T} > 0$ and $g_\text{gap} > 0$. (e) Experimental three-terminal $I-V$ characteristic of the contact for $V_\text{g} = 40\ \text{V}$ and fit to the model (red solid line). The green dashed line shows the calculated tunneling $I-V$ with $g_\text{gap} = 0$. (f) Fit (solid lines) to the experimental three-terminal $I-V$ characteristics (circles) at different gate voltages. The A and B parameters (see equation \ref{eqn3}), as well as $V_\text{th}$ are kept constant for the full set of $I-V$s. $V_\mathrm{g}$ is matched with the experimental applied voltages and $g_\text{gap}$ is used as the only fitting parameter. Inset: $g_\text{gap}$, extracted from the fittings as a function of the gate voltage. (g)Product of contact resistance and area ($R_\mathrm{C}.A$) as a function of $V_\mathrm{{g}}$, measured at different SD biases. The inset shows modulation of $R_\mathrm{C}.A$ in the zoomed-in range of ${V_\mathrm{g}}$.}
    \label{fig:four}
\end{figure*}

As shown, the performance of the TMD-based FETs is largely affected by the contact regions. We model the electrical behavior of the contacts using the band diagram of the metal/h-BN/TMD heterostructure shown in Figure \ref{fig:four}. The system can be conceptually understood as a parallel-plate capacitor with a small leakage current (tunneling through the h-BN). When the applied voltage between the metal and MoSe$_2$ is zero (Figure \ref{fig:four}a), the Fermi levels at the metal and MoSe$_2$ are the same, $E_\mathrm{F,M} = E_\mathrm{F,S}$. The band alignment between h-BN and MoSe$_2$ will be determined by the difference between their electron affinities. At the metal side, the band alignment will be given by the difference between the electron affinity of h-BN ($\rchi_\mathrm{BN}$) and the work function of titanium ($\phi_{\mathrm{M}}$). We consider $\phi_{\mathrm{M}}$ shifted by 0.78 eV due to the chemisorption of h-BN to Ti \cite{bokdam2014schottky}. The position of $E_\mathrm{F,S}$ with respect to $E_\mathrm{C}$ is controlled by the gate voltage. Importantly, the matching of the Fermi energies between Ti and MoSe$_2$ at equilibrium requires a misalignment of the vacuum levels at the two sides of the h-BN barrier. Thus, even at zero applied bias, a nonzero electrostatic voltage drop $V_\mathrm{BN}$ will appear at the h-BN layer.

$\mathrm{V_{3T}}$ is the voltage drop across the tunnel barrier which is the difference between the Fermi energies of the metal and semiconductor. When $V_\mathrm{3T}\neq0$ (see Figure \ref{fig:four}b-d), the Fermi energies at both sides become misaligned by $E_\mathrm{F,M}-E_\mathrm{F,S}=qV_{3T}$ where $q$ is the electron charge. Then, owing to the capacitive coupling between metal and MoSe$_2$, the charge density of MoSe$_2$ changes. Due to the limited density of available states, this will cause a shift in the position of the edge of the conduction band with respect to $E_\mathrm{F,S}$. From the band diagram of Figure \ref{fig:four}b, one gets the relation
\begin{equation}
    qV_\mathrm{BN}-qV_\mathrm{3T} = \phi_\mathrm{M}-\rchi_{\mathrm{MoSe}_2}-(E_\mathrm{F,S}-E_\mathrm{C})
    \label{eqn1}
\end{equation}

In order to clarify the meaning of equation \ref{eqn1}, we now consider two situations. First, let us assume that the density of states at the MoSe$_2$ near $E_\mathrm{F,S}$ is large (e.g., when the Fermi energy is well inside the conduction band of MoSe$_2$, Figure \ref{fig:four}b). Then, when the three-terminal voltage is increased by $\delta V_\mathrm{3T}$, charge carriers can accumulate in the MoSe$_2$ layer by filling empty states without noticeably shifting $E_\mathrm{F,S}$ with respect to $E_\mathrm{c}$, $\Delta(E_\mathrm{F,S}-E_\mathrm{C}) \simeq 0$ and, from equation \ref{eqn1}, $\Delta V_\mathrm{BN} \simeq \Delta V_\mathrm{3T}$. Thus, the shape of the h-BN tunnel barrier, as well as the energy window at which electrons can tunnel through the barrier (highlighted in yellow in figure \ref{fig:four}b) are strongly dependent on $V_\mathrm{3T}$. On the other hand, if the density of states at the MoSe$_2$ near the Fermi energy is very low (i.e. when $E_\mathrm{F,S}$ is inside the bandgap, Figure \ref{fig:four}c,d), charge can only be accumulated (or removed) by largely shifting $E_\mathrm{F,S}$ with respect to $E_\mathrm{c}$, $|\Delta(E_\mathrm{F,S}-E_\mathrm{C})| >> 0$ and therefore, $|\Delta V_\mathrm{BN}|<<|\Delta V_\mathrm{3T}|$. In this case, the shape of the h-BN barrier and the tunneling energy window are barely affected by changing V$_\mathrm{3T}$. Thus, the tunneling current through the barrier becomes saturated.
\begin{figure*}
    \centering
    \includegraphics[width=\textwidth]{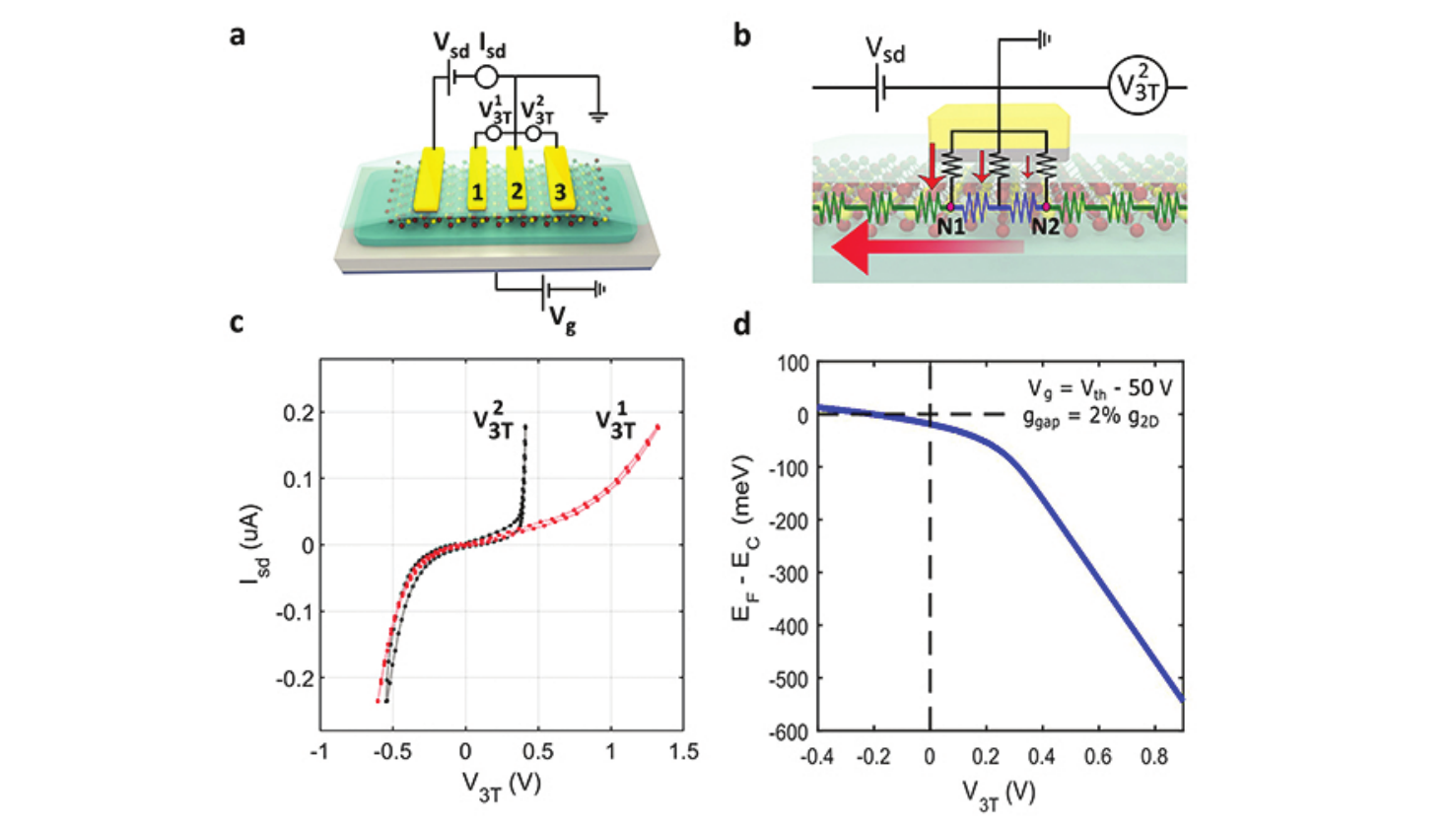}
   \caption{\textbf{Effect of current crowding on three-terminal measurements.} (a) Sketch of the device and 3T measurement geometries. (b) Resistor network at the metal/h-BN/TMD interface and the TMD channel. Red arrows depict the electron current and illustrate current crowding at the contact region. (c) Comparison of $V_\mathrm{{3T}}^1$ and $V_\mathrm{{3T}}^2$, measured as a function of $I_\mathrm{{sd}}$ at $V_\mathrm{{g}}= +50\ \text{V}$ in the geometry shown in panel (a). (d) Modulation of Fermi energy of MoSe$_\text{2}$ as a function of the 3T voltage, assuming a nonzero density of gap states and a back gate voltage of $V_\mathrm{g}=V_\text{th}-50 \text{V}$ in the model.
}
    \label{fig:five}
\end{figure*}

Next, we use the band diagrams discussed above to model the tunneling $I-V$ characteristic of the h-BN barrier. The tunneling current density, $J_\mathrm{tunnel}$ is approximately given by \cite{ponomarenko2013field}
\begin{equation}
\begin{split}
J_\mathrm{tunnel} = \mathrm{A} \times \sum_\mathrm{all\ bands}2 \int g_\mathrm{2D} \times P_\mathrm{tunnel}(E) \times\\
\left(f(E-E_\mathrm{F})-f(E-E_\mathrm{F}-V_\mathrm{3T})\right)dE\\
\end{split}
\label{eqn2}
\end{equation}
where we model the $\mathrm{MoSe_2}$ density of states as that of a 2D electron gas, $g_\mathrm{2D}=m_e^*/(\pi\hbar^2)$. $P_\mathrm{tunnel}$ is the transmission probability through the h-BN barrier, $f$ is the Fermi-Dirac distribution function and the sum runs for all the spin-orbit split subbands. A is a contact-dependent fitting parameter to be empirically determined. The factor 2 before the integrals accounts for the valley degeneracy.
For simplicity, we approximate the trapezoidal tunnel barrier (indicated in green in the band diagrams of figure \ref{fig:four}) by a rectangular barrier with an effective barrier height equal to the average height along the h-BN (orange line in the diagrams). We further assume that $P_\mathrm{tunnel}$ is exponentially dependent on the energy of the tunneling electrons as
\begin{equation}
\noindent
\begin{adjustbox}{max width=230pt}
$
P_\mathrm{tunnel}(E)=\exp\left( \frac{-\mathrm{B}\sqrt{2m^*}}{\hbar} \times\sqrt{U_\mathrm{barrier}-E} \right) 
$
\end{adjustbox}
\label{eqn3}
\end{equation}
where $U_\mathrm{barrier}=\rchi_{\mathrm{MoSe}_2}-\rchi_{\mathrm{BN}}-\frac{1}{2} qV_\mathrm{BN}$ is the tunneling barrier height and $E$ is the electron energy with respect to $E_\mathrm{C}$. B, as A in equation \ref{eqn2}, is used as a contact-dependent fitting parameter. Note that, for the case of isotropic, parabolic bands one gets B = 1, while we get the best fit to our experimental data considering B=0.3. Figure \ref{fig:four}e shows the fit (red line) to an experimental three-terminal $I-V$ (markers) with $V_\mathrm{gate} = 40\ \text{V}$. The $I-V$ characteristic can be divided in two different regimes, depending on whether the $E_\mathrm{F,S}$ is above (figures \ref{fig:four}a, b) or below (\ref{fig:four}c, d)  $E_\mathrm{C}$. For $E_\mathrm{F,S} > E_\mathrm{C}$, the density of states at the $\mathrm{MoSe_2}$ is large. Thus, the barrier height $U_\mathrm{barrier} \propto V_\mathrm{BN}$ will change almost linearly with $V_\mathrm{3T}$. For $E_\mathrm{F,S} < E_\mathrm{C}$, however, the density of states becomes much lower. Thus, $U_\mathrm{barrier}$ changes more slowly with $V_\mathrm{3T}$. The transition between these two regimes creates a kink in the $I-V$ curves at $V_\mathrm{3T}	\approx 0.3 \text{V} $. In the ideal case where the density of gap states $g_\mathrm{gap}$ is zero, $U_\mathrm{barrier}$ remains constant, and the tunneling current becomes completely saturated (green, dashed line in figure \ref{fig:four}e). In order to reproduce the experimental $I-V$ we need to consider a nonzero density of gap states $g_\mathrm{gap}$. For simplicity, we consider $g_\mathrm{gap}$ to be energy-independent (see SI, section 6) $g_\mathrm{gap}=1.8\%\ g_\mathrm{2D}$, which leads to a nonideal current saturation (red, solid line). It is worth noting that in reality, $g_\mathrm{gap}$ is expected to be markedly energy-dependent, with a large density of gap states concentrated at energies near the conduction and valence band edges (further discussed in SI). An additional fitting parameter is the threshold gate voltage $V_\mathrm{th}$, which shifts the position of $E_\mathrm{F,S}$ with respect to $E_\mathrm{C}$ and thus determines the value of $V_\mathrm{3T}$ at which $E_\mathrm{F,S} = E_\mathrm{C}$. From the fit to the experimental $I-V$ we get $V_\mathrm{th} = 90\ \text{V}$, well above the values extracted from four-terminal transfer characteristics. This indicates that the $\mathrm{MoSe_2}$ flake is depleted by the proximity of the metallic contacts, as also discussed in section 3.

Figure \ref{fig:four}f shows the experimental three-terminal $I-V$ characteristics (markers) for different gate voltages $V_\text{g}$ ranging from 30 to 60$\ \text{V}$. The nonlinear 3T $I-V$ curves show the tunneling diode behavior of the contact with forward negative bias. We fit the full set of I-Vs to equation \ref{eqn3} (solid lines), using $g_\text{gap}$ as the only free parameter, while keeping A, B and $V_\mathrm{th}$ fixed. The full set of curves is well reproduced by the model considering a small density of gap states. The inset panel in Figure \ref{fig:four}f shows the values of $g_\text{gap}$ extracted from the fits, increasing from $g_\text{gap} = 0.8\% g_\text{2D}$ at $V_\text{g} = 30\ \text{V}$ to $g_\text{gap} = 6\% g_\text{2D}$ at $V_\text{g} = 60\ \text{V}$. This increase of $g_\mathrm{gap}$ for large gates voltages suggests that the density of gap states is indeed energy-dependent (rather than constant, as considered in the model, see SI), with a larger concentration of states for energies close to the conduction band edge. In order to check the consistency of the values of $g_\text{gap}$ extracted from the fits, we estimate its lower limit from the four-terminal transfer curves (with a $V_\text{g}$ sweep of $\pm 60\ \text{V}$ and $V_\text{th}= +20\ \text{V}$). Assuming that switching from conduction to valence band requires at least a $V_\text{gate}$ shift of 80$\ \text{V}$, we get $g_\mathrm{gap} \approx 1.5\%\ g_\text{2D}$, compatible with the values extracted from the model.

Figure \ref{fig:four}g shows the contact resistance-area product ($R_\mathrm{C}.A$) as a function of $V_\mathrm{g}$. The contact resistance ($R_\mathrm{C}$) is defined as a combination of the metal/SC interface resistance and the resistance of the SC channel beneath the contact \cite{cohen1982contact}. We observe strong modulation of $R_\mathrm{C}.A$ as a function of $V_\mathrm{{g}}$ for the range of $V_\mathrm{{g}} < 40\ \text{V}$ at high $V_\mathrm{{sd}}$, which is not as large for low $V_\mathrm{{sd}}$. Consistent with our explanation regarding the reverse (positive) bias  of the Ti/h-BN/TMD tunneling diode, the positive $V_\mathrm{{sd}}$ lowers the $E_\mathrm{F}$ towards the SC bandgap, where changing the gate voltage can considerably tune the density of the sates. However, at very low $V_\mathrm{{sd}}$, the $E_\mathrm{F}$ is still positioned in or close to the conduction band. Therefore the change in the density of states and so the contact resistance as a function of $V_\mathrm{{g}}$ is not as dramatic. The strong gate and bias dependence of the contact resistance and the tunneling current confirms efficient gating of the TMD channel beneath the contacts.

This considerable gating effect of the bias applied on the contact is further evidenced by comparison of the two three-terminal (3T) measurement geometries, shown in Figure \ref{fig:five}a. We also illustrate the resistor network of the metal/h-BN/TMD interface (in Figure \ref{fig:five}b), consisting of metal/h-BN/TMD contact resistance (in black), TMD channel beneath the contact (in blue) and TMD channel in between the two contacts (in green). In the 3T geometries, $V_\mathrm{{3T}}^1$ is the summation of the voltage drop on contact 2 (voltage on node N1 of Figure \ref{fig:five}b) and the voltage drop across the channel in between contact 1 and 2. The $V_\mathrm{{3T}}^2$ in this geometry measures the voltage of node N2 (of contact 2). 

Measurement of $V_\mathrm{{3T}}^1$ and $V_\mathrm{{3T}}^2$ as a function of $I_\mathrm{{sd}}$ is compared in Figure \ref{fig:five}c. As discussed, the channel resistance is order(s) of magnitude smaller than contact resistances. Therefore, the $I$-$V$ characteristics from Figure \ref{fig:five}c are dominated by the voltage drop on the contact (C2) in both geometries. However, for almost all of the crossing contacts we observe distinct behavior of $V_\mathrm{{3T}}^1$ and $V_\mathrm{{3T}}^2$ for the positive range of the 3T voltages. Further, the deviation exactly occurs at the kink mentioned in the 3T $I-V$ of Figure \ref{fig:four}e, where the Fermi energy is below $E_\mathrm{{C}}$. Such charge depletion underneath the contact makes the in-plane resistivity (blue resistors of Figure \ref{fig:five}b) to be comparable to the out-of-plane resistivity (black resistors). In this case the charge current that flows through the left side of the contact (node N1) is higher than that of the right side (node N2) and therefore the $V_\mathrm{{3T}}^2$ saturates to lower values than $ V_\mathrm{{3T}}^1$.

Figure \ref{fig:five}d shows the shift of $E_\textrm{F}$ with respect to $E_\textrm{C}$ as a function of the three-terminal voltage for $V_\mathrm{g} = V_\textrm{th}-50\ \text{V} = 40\ \text{V}$. The kink in this diagram corresponds to the point at which the conduction band becomes fully depleted, closely matching the voltage at which the $I$-$V$s from panel c start to diverge. Note that, as discussed above, even relatively small changes in $V_\textrm{3T}$ can strongly modulate the doping of the channel. In particular, we observe that, even if the $\mathrm{SiO_2}$ back gate is 50 V below the threshold voltage, a $V_\textrm{3T}$ as small as -0.2 V is sufficient for $E_\textrm{F}$ to reach the edge of the conduction band. 

\section{Conclusions}

Our results show that metal/bilayer h-BN electrodes are very promising for low-energy transport in 1L TMD-based devices, allowing to reduce the effects of Fermi level pinning and formation of Schottky barriers. In consequence, we find that bilayer h-BN tunnel contacts outperform direct metal/SC contacts both in terms of electrical response and quality, yielding reduced hysteresis. Elimination of metal/TMD chemical interactions by the h-BN insertion layer along with the full BN-encapsulation of the channel and the side electrode geometry, allows to better preserve the intrinsic properties of the 2D channel and reach carrier mobilities comparable to those of the pristine $\mathrm{MoSe_2}$ even at room temperature. The model described here for h-BN tunnel contacts to 2D semiconductors allows to satisfactorily describe and reproduce their experimental electrical response, showing that the doping of the semiconductor channel below the contacts can be largely modified even for relatively small source-drain bias voltages.

\section{Methods}
Atomically thin layers of $\mathrm{MoSe_2}$ and h-BN are mechanically cleaved from their bulk crystals on $\mathrm{SiO_2}$/Si substrates, using adhesive tapes \cite{novoselov2005two}. We identify the thin flakes by their optical contrast with respect to the substrate \cite{li2013rapid,benameur2011visibility} and verify their thicknesses by atomic force microscopy (AFM). Using dry pick-up technique \cite{Zomer}, we pick up the bilayer h-BN flake by PC (Poly(Bisphenol A)carbonate) and  a PDMS stamp. We use the bilayer h-BN flake on the PC layer to pick up monolayer $\mathrm{MoSe_2}$ by means of vdW interactions between the h-BN and $\mathrm{MoSe_2}$ flakes. Then, we release the picked-up flakes on top of the bulk h-BN on $\mathrm{SiO_2}$ (300 nm)/doped Si substrate, by melting the PC. The PC layer is dissolved in Chloroform for 5 min and the residues are removed by annealing the sample in Ar/$\mathrm{H_2}$ flow at 350 °C for 3 hours. We proceed with fabrication of electrodes on the vdW stack by e-beam lithography technique (using PMMA as the e-beam resist) and e-beam evaporation of Ti (5nm)/Au (75 nm) at the pressure of $10^{-6}$ mbar, followed with lift-off in Acetone at $40\mathrm{^{\circ}C}$. The UHV condition for evaporation of the electrodes is to minimize interfacial contamination and more importantly to avoid oxidation of Ti. 
 
\section{Acknowledgements}
We kindly acknowledge H. M. de Roosz, T. J. Schouten, H. Adema and J. G. Holstein for technical support. This research has received funding from the Dutch Foundation for Fundamental Research on Matter (FOM) as a part of the Netherlands Organisation for Scientific Research (NWO), FLAG-ERA (15FLAG01-2), the European Union’s Horizon 2020 research and innovation programme under grant agreements No 696656 and 785219 (Graphene Flagship Core 1 and Core 2), NanoNed, the Zernike Institute for Advanced Materials, and the Spinoza Prize awarded to B. J. van Wees by NWO.

\section{References}
\bibliography{bibl}

\pagebreak
\onecolumn
\renewcommand{\figurename}{Figure S}
\setcounter{figure}{0} 

\title[ ]{Supporting Information\\Bilayer h-BN Barriers for Tunneling Contacts in Fully-Encapsulated Monolayer $\mathbf{MoSe_2}$ Field-Effect Transistors}
\author{Talieh S. Ghiasi$^1$, Jorge Quereda$^1$, Bart J. van Wees$^1$}

\address{$^1$Zernike Institute for Advanced Materials, University of Groningen, Groningen, 9747 AG, The Netherlands}

\ead{t.s.ghiasi@rug.nl}

\newpage

\textbf{Contents}\\

1. Device Fabrication\\

2. AFM Characterization\\

3. Electrical performance of non-encapsulated and BN-encapsulated 
   $\mathrm{MoSe_2}$ FETs\\
   
4. Four-terminal measurements\\

5. Three-terminal measurements\\

6. Modeling of Metal/h-BN/TMD Contacts\\ 

7. References\\

\newpage
\textbf{1. Device Fabrication}\\

The device fabrication process is explained in the Methods section of the main manuscript in details. Here, we show the optical microscope (OM) image of the top layer h-BN flake and that of 1L-$\mathrm{MoSe_2}$ on 90 nm and 300 nm $\mathrm{SiO_2}$ substrates, respectively (Figure S\ref{fig:one}a and b). The top h-BN layer is picked up with the PC (Poly(Bisphenol A)carbonate) stamp and is used to transfer the $\mathrm{MoSe_2}$ flake on top of the bulk h-BN, where we release the flakes by melting the PC. Figure S\ref{fig:one}c shows the OM image of the vdW stack of BN-encapsulated $\mathrm{MoSe_2}$ on the $\mathrm{SiO_2/Si}$ substrate. The fabricated device by e-beam lithography of the Ti/Au electrodes is shown in Figure S\ref{fig:one}d. 

\begin{figure*}[ht]
\centering
\includegraphics[width=400pt]{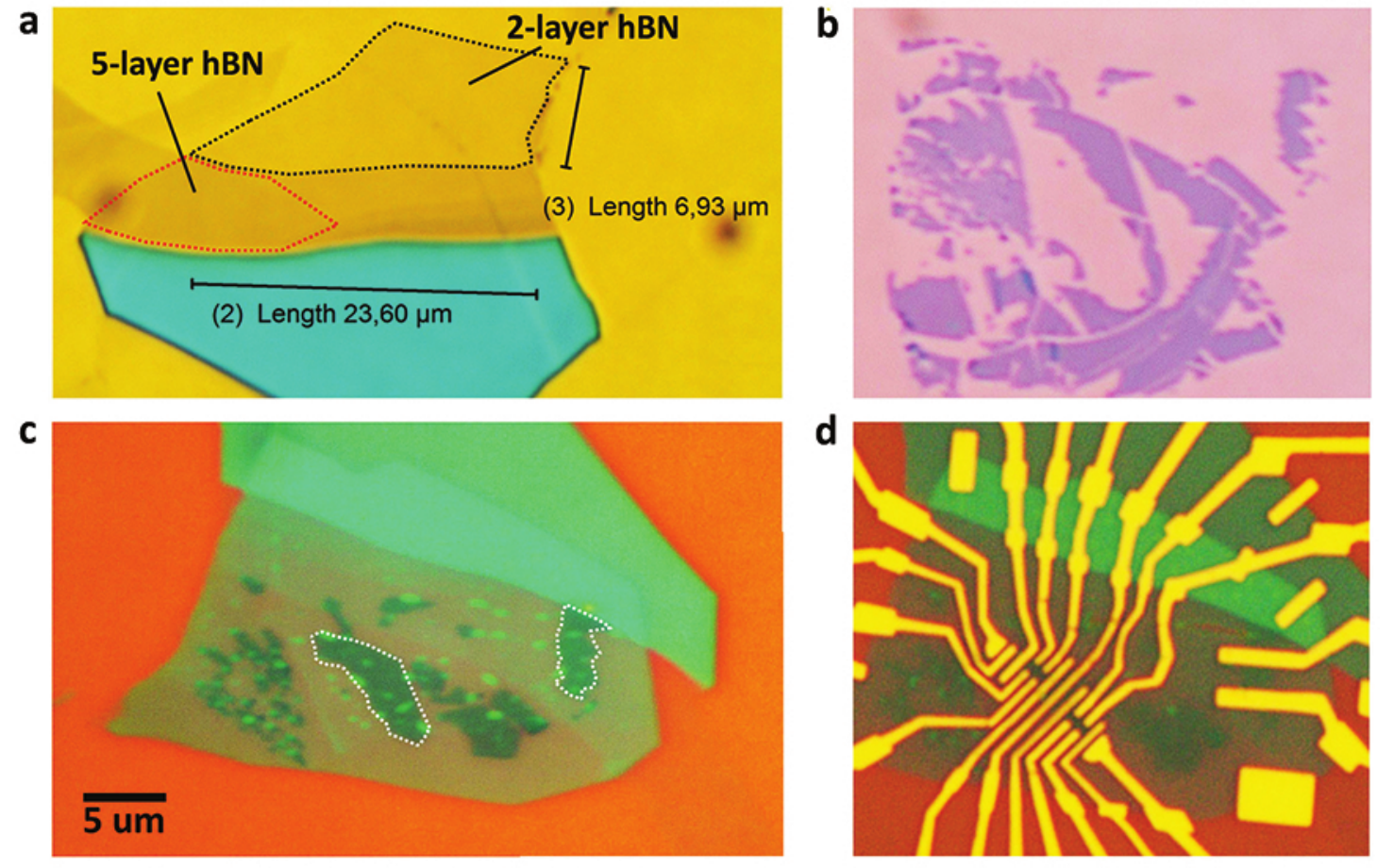}
\caption{\textbf{Fabrication of BN-encapsulated monolayer $\mathbf{MoSe_2}$ FET.} (a) OM image of the h-BN flake on 90 nm $\mathrm{SiO_2}$ substrate. This flake is used as a top layer in our vdW stack. The regions of 2-layer and 5-layer h-BN are determined by the black and red dashed lines, respectively. (b) OM image of the monolayer $\mathrm{MoSe_2}$ on 300 nm $\mathrm{SiO_2}$ substrate. (c) OM image of the vdW stack of the BN-encapsulated $\mathrm{MoSe_2}$, after removing PC. The white dashed lines show the edges of the encapsulated $\mathrm{MoSe_2}$ flakes that are used as the 2D channel for the FETs. (d) OM image of the fabricated device with Ti/Au electrodes.}
\label{fig:one}
\end{figure*}

\newpage
\textbf{2. AFM characterization}\\

We obtain the height profile of the vdW stack of the BN-encapsulated $\mathrm{MoSe_2}$ by atomic force microscopy (AFM) measurements (Figure S\ref{fig:two}a). The thickness of the $\mathrm{MoSe_2}$ and the top layers of h-BN are determined with respect to the bulk h-BN (Figure S\ref{fig:two}b). The thickness of the $\mathrm{MoSe_2}$ flakes is about 0.7 nm which corresponds to a monolayer of this material \cite{benameur2011visibility}. These flakes are shown with the black dashed-lines in the AFM image. The top h-BN flake has two regions that are covering the $\mathrm{MoSe_2}$ flakes. These two regions have the thickness of 0.7 nm and 1.85 nm that corresponds to the thickness of 2-layer and 5-layer of h-BN crystals \cite{golla2013optical} (edges of these two regions are highlighted with the green and yellow dashed-lines in the AFM image). 
The bulk h-BN that is used as the bottom layer, has the thickness of 7.65 nm (21-22 layers), measured on the $\mathrm{SiO_2}$ substrate.
The light spots in the AFM image, that is also observable in the OM image of Figure S\ref{fig:one}c, are related to the bubbles formed at the interfaces between the 2D materials.  
\begin{figure*}[ht]
\centering\includegraphics[width=\textwidth]{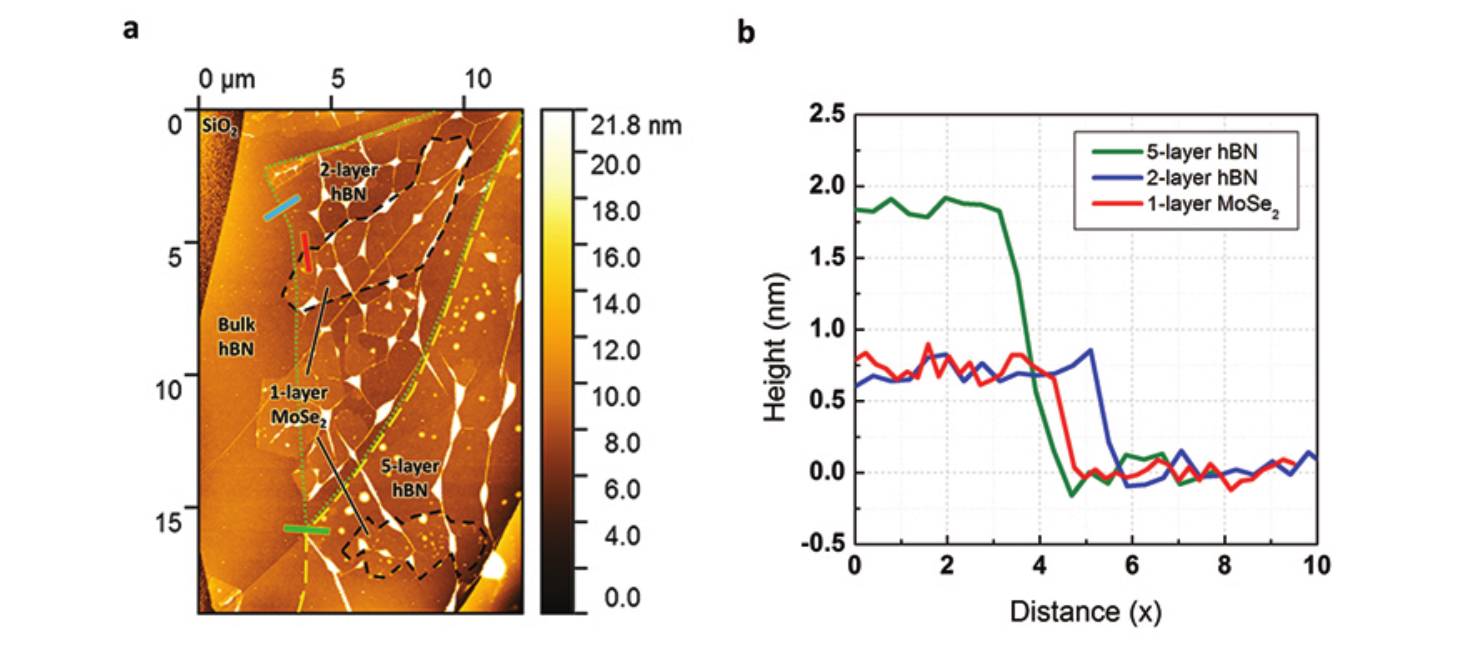}
\caption{\textbf{AFM characterization of the BN-encapsulated $\mathbf{MoSe_2}$ vdW heterostructure.} (a) AFM image of the vdW stack. the black dashed lines show the edge of the monolayer $\mathrm{MoSe_2}$ flakes. The green and yellow dashed lines show the edges of the 2-layer and 5-layer h-BN flakes, respectively. (b) Height profile of the $\mathrm{MoSe_2}$ flake and the two regions of the top h-BN flake (shown in Figure \ref{fig:one}a).}
\label{fig:two}
\end{figure*}

\newpage
\textbf{3. Electrical performance of non-encapsulated and BN-encapsulated $\mathbf{MoSe_2}$ FETs}\\

\begin{figure*}[ht]
\centering\includegraphics[width=9cm]{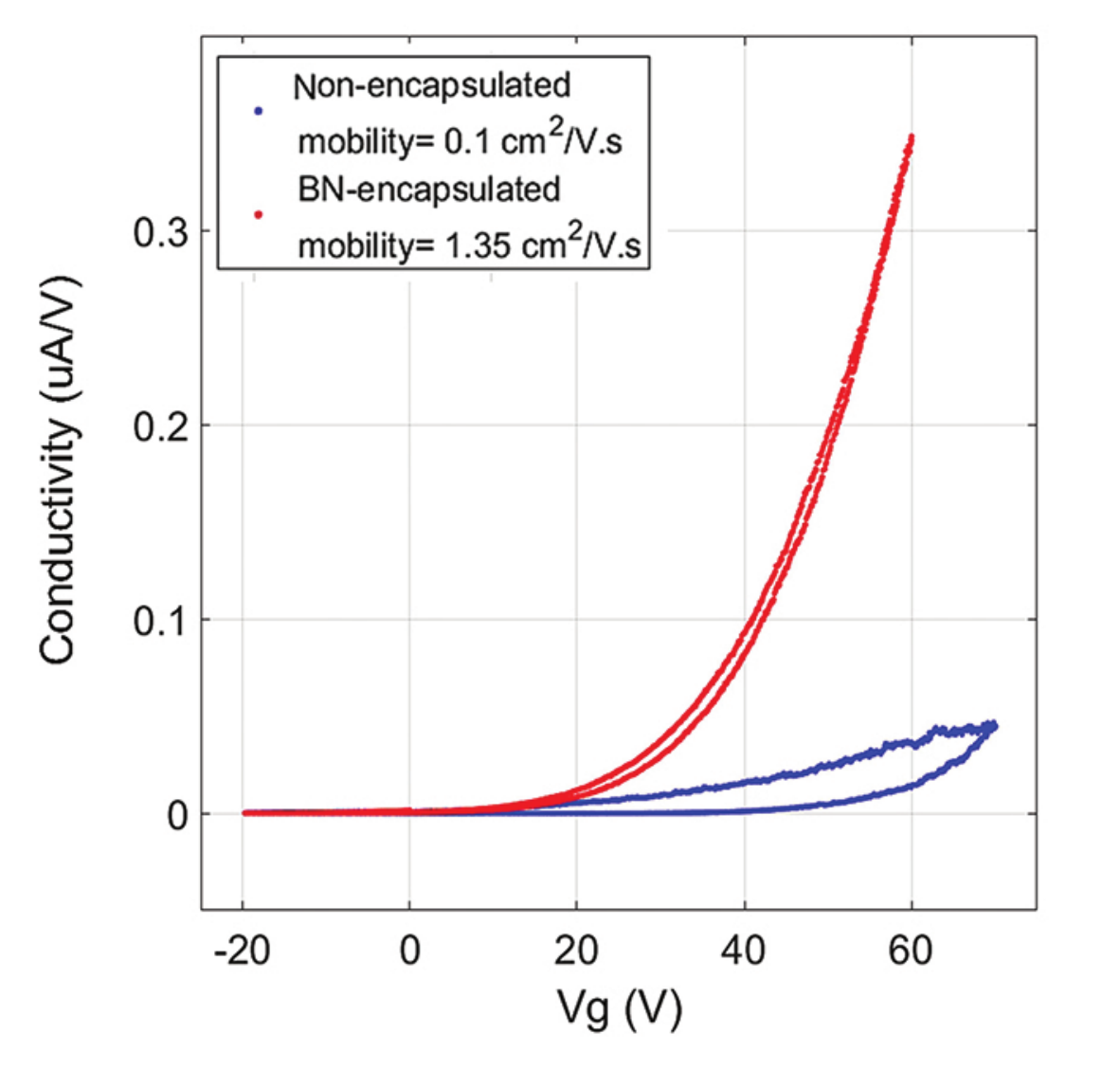}
\caption{Comparison of two-terminal measurements, performed on the FETs made with BN-encapsulated and non-encapsulated monolayer $\mathrm{MoSe_2}$ channel.}
\label{fig:seven}
\end{figure*}

To show the importance of BN-encapsulation on the performance of TMD-based FETs, we compare the electrical transport in the BN-encapsulated and non-encapsulated samples. In the non-encapsulated FET, the monolayer $\mathrm{MoSe_2}$ channel is in direct contact with Ti electrode and the Si substrate. In Figure S\ref{fig:seven}, we show the gate dependence of the conductivity, measured in two-terminal geometry with the gate sweep rate of 1 V/s. 

We observe a considerable increase in the two-terminal conductivity and mobility by BN-encapsulation of the channel. As discussed in the manuscript, two-terminal measurements on monolayer TMDs are dominated by the electrical behavior of the contacts. Therefore the increase in the two-terminal conductivity in the BN-encapsulated FET is not only due to reduction of roughness and Coulomb scattering in the channel, but also a decrease in the contact resistances by using bilayer h-BN as an insertion layer at metal/TMD interface. This decrease in the contact resistance is because of the lower Schottky barrier and reduction of the metal workfunction\cite{bokdam2014schottky} by using bilayer h-BN at the metal/TMD interface. This modulation of the metal work function by h-BN cause a negative shift in the threshold voltage (compared with the non-encapsulated sample in Figure S\ref{fig:seven}), consistent with recent observations\cite{wang2016high}.       
 
\newpage
\textbf{4. Four-terminal measurements}\\

We show that in the four-terminal (4T) measurements performed with the side contacts (shown in device sketches of Figure S\ref{fig:three}), the distance of the source-drain (SD) electrodes to the side 4T voltage probes matters. We observe that the measurements performed with the SD electrodes, closest to the voltage probes, show unconventional non-linear behavior in the $I-V$ curves (Figure S\ref{fig:three}a), while the ones measured with the SD electrodes that are further away from the inner probes show linear $I-V$ behavior (Figure S\ref{fig:three}b). We attribute this difference to the fact that the SD voltage can effectively gate the SC channel and tune the density of charge carriers close to the contact regions due to the presence of the 2L h-BN. This modulation of density of states close to the SD contacts can affect the voltage drop measured by the inner probes if the distance between them is not large enough. The results shown in the main manuscript (Figure 3b and d) corresponds to the measurement geometry of Figure S\ref{fig:three}b. 
  
\begin{figure*}[ht]
\centering\includegraphics[width=0.95\textwidth]{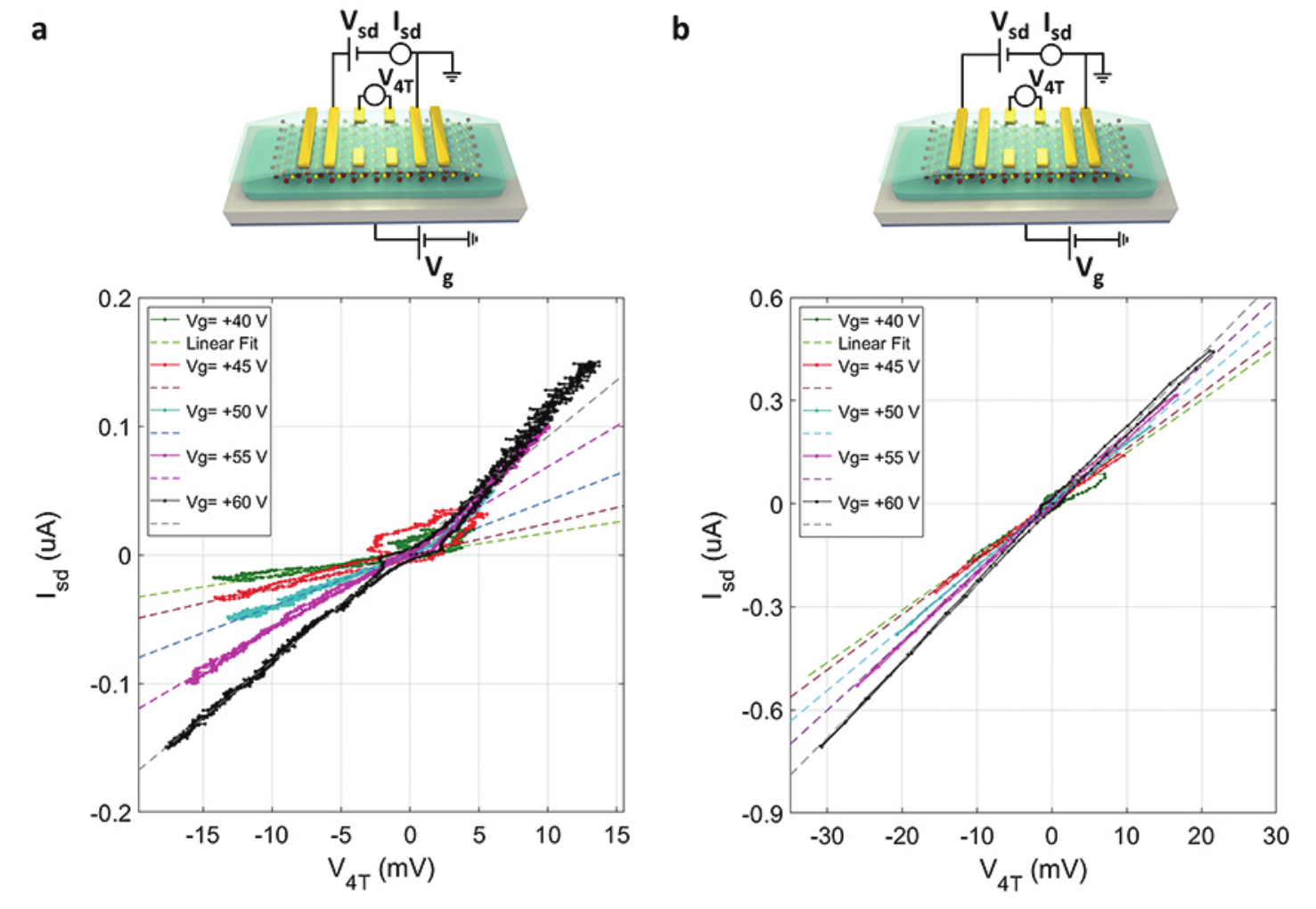}
\caption{\textbf{Four-terminal measurements.} The measurements are performed with the SD bias applied to the outer contacts, while the 4T voltage is measured with the inner side electrodes. Sketches of the contact geometry and electrical circuits are illustrated.(a) The 4T $I-V$ measurements, performed with the outer SD electrodes close to the inner voltage probes with $0.7 \mu m$ spacing. The dashed lines are the linear fits only to the negative range of $I-V$s. (b) The 4T $I-V$ measurements, performed with the outer SD electrodes that are further away from the inner voltage probes with $1.6 \mu m$ spacing.}
\label{fig:three}
\end{figure*}

\newpage
\textbf{5. Three-terminal Measurements}\\

In TMD-based FETs (specifically with direct contact of metal/TMD) it is observed that due to the presence of highly depleted TMD channel underneath the contacts, the charge injection/detection is preferably happening through the edges of the contacts. This can be mainly problematic in the 4T or 3T measurements (i.e. $\mathrm{V_{3T}}^1$ in Figure S\ref{fig:five}a), in which the voltage probe is placed between the source and drain electrodes. In this case, as shown in Figure S\ref{fig:five}a, the SD current can flow through the middle contact to avoid the highly depleted region underneath the contacts.  

Here we investigate the possibility of current flow through the metal/h-BN/TMD contacts. We perform the 3T measurements in the configurations shown in Figure S\ref{fig:five}b. The 3T $I-V$ curves of Figure S\ref{fig:five}c show the distinct behavior of $\mathrm{V_{3T}}^1$ and $\mathrm{V_{3T}}^2$ as already explained, while $\mathrm{V_{3T}}^3$ shows exactly same behavior as of $\mathrm{V_{3T}}^1$. This agreement of $\mathrm{V_{3T}}^1$ and $\mathrm{V_{3T}}^3$ is a clear indication for no current flow through the crossing contact (probe of $\mathrm{V_{3T}}^1$), which is an important improvement in the electrical behavior of the contacts.    

\begin{figure*}[ht]
\centering\includegraphics[width=\textwidth]{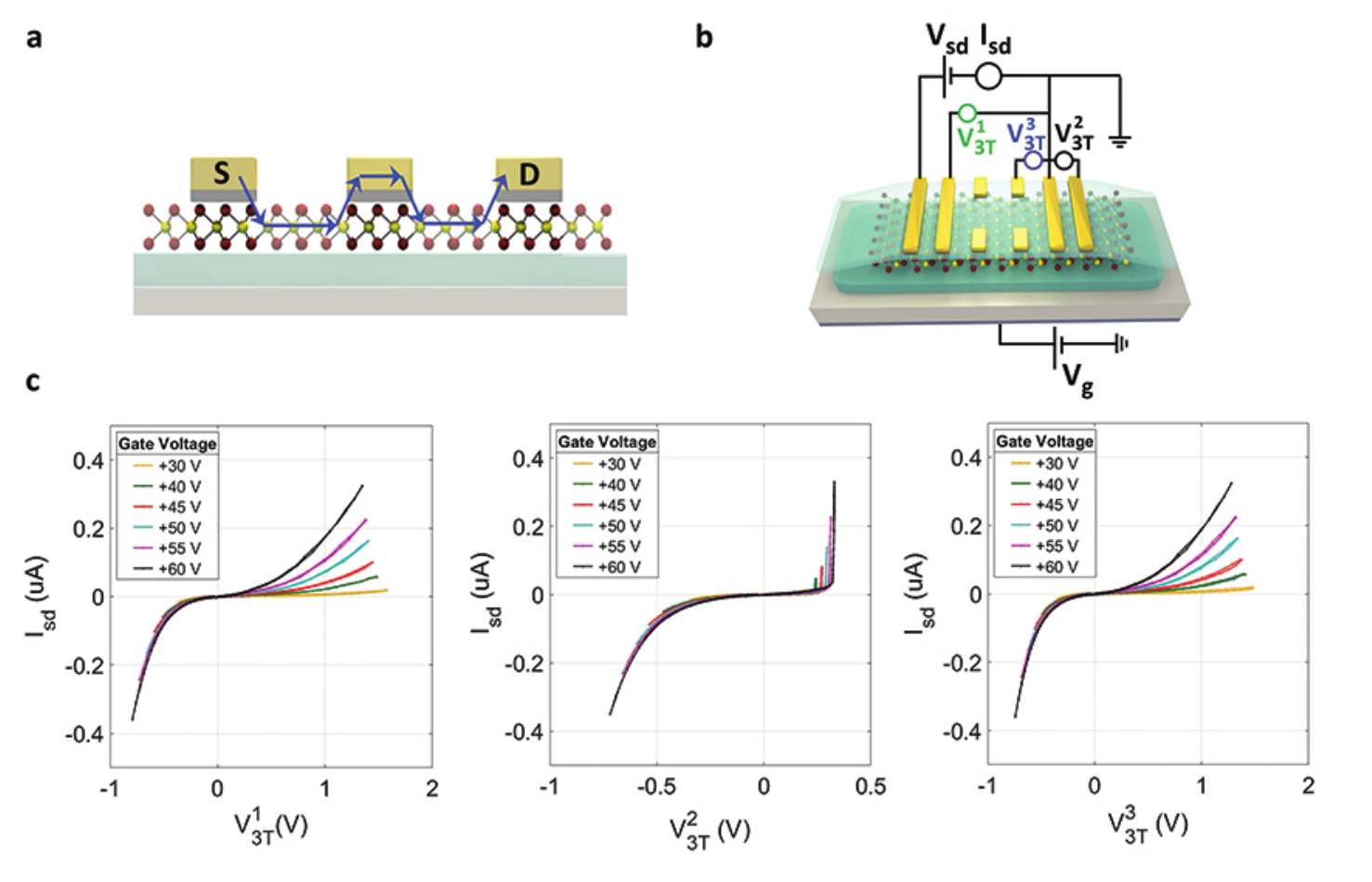}
\caption{\textbf{Comparison of three-terminal measurements.} (a) SD current injection/detection through the edges of SD contacts. Current flow through the middle contact, avoiding the depleted TMD channel underneath the contact. (b) three-terminal measurement geometries. (c) corresponding 3T $I-V$ characteristics}
\label{fig:five}
\end{figure*}

\newpage
\textbf{6. Modeling of Metal/h-BN/TMD Contacts}\\

We start by considering a three-plate capacitor. 
The first plate represents the metallic contact, the central plate represents the 2D semiconductor and the third plate represents the doped Si back gate. Each plate will cause an electric field in the dielectric regions given by

\begin{equation}
    \left|{\vec{E}}\right| = \frac{n q}{2\epsilon_\textrm{0}\epsilon_\textrm{d}}
\end{equation} 
where $\vec{E}$ is the electric field, $n$ is the number of charge carriers accumulated at the plate, $-q$ is the electron charge, $ \epsilon_\textrm{0} $ is the vacuum electric permittivity
and $\epsilon_\textrm{d}$  is the relative permittivity of the dielectric material. Therefore, the voltage drop, $V$ at the dielectric regions will be given by

\begin{equation}
qV_{\textrm{BN}}=\frac{d_\textrm{BN}q}{2\epsilon_\textrm{0}\epsilon_\textrm{BN}}\left( n_\textrm{M}-n_\textrm{TMD}-n_\textrm{Si}\right) 
\label{eq1}
\end{equation}

\begin{equation}
qV_{\textrm{SiO}_2}=\frac{d_{\mathrm{SiO_2}}q}{2\epsilon_\textrm{0}\epsilon_{\textrm{SiO}_2}}\left( n_\textrm{M}+n_\textrm{TMD}-n_\textrm{Si}\right)
\label{eq2}
\end{equation}
where the subindices $\textrm{BN}$ and ${\textrm{SiO}_2}$ correspond to the two different dielectric regions and $n_\textrm{M}$, $n_\textrm{TMD}$ and $n_{\textrm{SiO}_2}$ are the density of charge carriers the metallic contact, the 2D semiconductor and the doped silicon gate, respectively. We also require that the field outside the capacitor is zero, which yields

\begin{equation}
n_\textrm{M}+n_\textrm{TMD}+n_\textrm{Si}=0 .
\label{eq3}
\end{equation}

Solving the set of equations \ref{eq1}, \ref{eq2} and \ref{eq3}  we get an expression for $n_{\textrm{TMD}}$ as a function of the electrostatic voltage drops:

\begin{equation}
n_{\textrm{TMD}}=-\epsilon_\textrm{0} \left( \frac{\epsilon_\textrm{BN}}{d_\textrm{BN}}V_\textrm{BN} + \frac{\epsilon_{\textrm{SiO}_2}}{d_{\textrm{SiO}_2}}V_{\textrm{SiO}_2} \right)
\label{eq4}
\end{equation}
Note that $\epsilon_{\mathrm{BN}}/{d_\mathrm{BN}}\approx200\times\epsilon_{\mathrm{SiO_2}}/{d_\mathrm{SiO_2}}$. Thus, the top contact is 200 times more effective than the back gate in tuning $n_{\textrm{TMD}}$. 

The carrier density at the semiconductor can also be written as
\begin{equation}
n_{\textrm{TMD}}=\int_{-\infty}^{\infty} g_{\textrm{TMD}}(E)f(E-E_\textrm{F})dE ,
\label{eq5}
\end{equation}
where $g_{\textrm{TMD}}$ is the density of states of the TMD and $f(E-E_\textrm{F})$ is the Fermi-Dirac distribution function centered at $E_\textrm{F}$. We now assume that the density of states in each spin-orbit split subband of the TMD is that of a two-dimensional electron gas, $g_{\textrm{2D}}=m_e^*/\pi\hbar^2$. In order to take into account the presence of localized states in the bandgap due to impurities and interface states, we consider the density of states in the gap to be a small fraction $\alpha$ of $g_{\textrm{2D}}$:  $g_{\textrm{gap}}=\alpha g_{\textrm{2D}}$. Then, the integral in equation \ref{eq5} can be split into the contributions from each subband and from the gap states:

\begin{equation}
\begin{split}
n_{\textrm{TMD}} = n_\textrm{0} + g_\textrm{2D}
&\left(\int_{0}^{\infty} f(E-E_\textrm{F})dE  
+ \int_{E_\textrm{SO}^\textrm{CB}}^{\infty}f(E-E_\textrm{F})dE
+ \alpha \int_{-E_\textrm{g}}^{0}f(E-E_\textrm{F})dE \right.
\\
&\left. 
- \int_{-\infty}^{-E_\textrm{g}} \left(1-f(E-E_\textrm{F})\right)dE
- \int_{-\infty}^{-E_\textrm{g}-E_\textrm{SO}^\textrm{VB}} \left(1-f(E-E_\textrm{F})\right)dE
\right) ,
\label{eq6}
\end{split}
\end{equation}
where we have fixed the energy origin at the edge of the conduction band, so that
$E_\textrm{C}=0$ and $n_{\textrm{TMD}}(E=0)=n_\textrm{0}$. Finally, by matching the right sides
of equations \ref{eq4} and \ref{eq6} we can obtain a relation between the applied voltages
$V_{\textrm{SiO}_2}$, $V_{\textrm{BN}}$ and the Fermi energy $E_\textrm{F}$, that must be solved
numerically to get $E_\textrm{F}(V_\textrm{BN},V_{\textrm{SiO}_\textrm{2}})$.
Once $E_\textrm{F}$ is calculated, one can follow the method discussed in the main text to calculate the $I-V$ characteristics of the tunnel contact.



\end{document}